\documentclass[osajnl,twocolumn,showpacs,superscriptaddress,10pt]{revtex4-1} 
\usepackage{amsmath,amssymb,graphicx}
\pdfoutput=1
\usepackage{array}
\usepackage[caption=false]{subfig}
\usepackage{bm}
\usepackage{amsfonts}
\usepackage{epsfig}

\def\be{\begin{equation}}
\def\ee{\end{equation}}
\def\ba{\begin{align}}
\def\nn{\nonumber\\}

\begin{document}

\title{\bf 3D Source Localization and Polarimetry using High Numerical Aperture Imaging
with Rotating PSF}
\author{Zhixian Yu} 
\author{Sudhakar Prasad} \email{Submitted to Optics Letters}
\affiliation{Department of Physics
and Astronomy, University of New Mexico, Albuquerque, New Mexico 87131}

\begin{abstract}
Rotating-PSF imaging via spiral phase engineering can localize point sources 
over large focal depths in a snapshot mode. 
This letter presents a full vector-field analysis of the rotating-PSF imager
that quantifies the PSF signature of the polarization state of the imaging light.
For sufficiently high image-space numerical apertures, there can be significant 
wave-polarization dependent contributions to the overall PSF, which would allow one to 
jointly localize and sense the polarization state of light emitted by point 
sources in a 3D field.

\end{abstract}

\maketitle
\vspace{-10pt}

The orbital angular momentum (OAM) of light can 
encode the axial position of a source point via the rotation of its image,
the so-called point spread function (PSF).
As one of us has shown \cite{SP13}, an annular spiral phase structure
in the circular pupil of an imager yields a PSF that, as a coherent superposition
of non-diffracting Bessel beams, rotates about the Gaussian image point nearly rigidly 
as the source is displaced axially from the Gaussian image plane (GIP).
In its superior depth of field and PSF compactness, 
this system vastly improves upon previous rotating-PSF imagers \cite{PP08,LLBM11,GDQDP12}, 
all of which combine fully diffracting Gauss-Laguerre vortex modes in the pupil. 
The angle of the PSF rotation, being 
proportional to the axial displacement, or defocus, of the source from the GIP, thus
enables one to localize a source fully in all three dimensions.

For rotating-PSF imaging with a high image-space numerical aperture (NA), 
we must include the transversality of the electromagnetic field
via a vector-field analysis \cite{MM92,JK06,THYH13}. Such an analysis accounts properly for wave polarization,
related fundamentally to the spin angular momentum (SAM) of the photon, and is thus needed to describe
polarization-dependent modifications of the rotating PSF. 
This is a subtle effect, as we shall see, one that is best seen via the nonvanishing longitudinal 
field components of the imaging beam \cite{Jackson99}.
One expects SAM to modify the rotating PSF amplitude from its OAM-only form by an amount proportional to the 
product of the NA and that contributed by one unit of angular momentum when the light is completely circularly polarized, 
with still subtler modifications for more general wave polarization states. 
It is this modification and its exploitation for
snapshot 3D polarimetric imaging that is the subject of the present letter.

The spiral pupil phase structure which yields a compact PSF that rotates without
much distortion is
\be
\label{e1}
\Psi(u,\phi_u)=\left\{f_l(\phi_u) \,\Bigg|\, \sqrt{l-1\over L}\leq u<\sqrt{l\over L},\ l=1,\ldots,L\right\},
\ee
where $u=\rho/R$ is the normalized radial coordinate in the circular pupil of radius $R$ and
$\phi_u$ is the azimuthal angle. The simplest form for the spiral phase $f_l(\phi_u)$
in the $l$th annular zone is, up to an unimportant additive constant, $\pm l\phi_u$ \cite{SP13}, 
for which a single-lobe PSF results. However, 
any spiral phase distribution with an integral winding number in each zone that changes by a fixed step
from one zone to the next will do, with the number of lobes of the resulting PSF being equal to the step size
of the phase winding number. 


Consider a point electric dipole source located at the origin, with dipole moment $\vec p$ oscillating at angular 
frequency $\omega$ in a plane transverse to the $z$ axis,
\be
\label{e2}
\vec p(t) =(p_+\hat e_+ + p_-\hat e_-)\exp(-i\omega t),
\ee
where $\hat e_\pm$ are the circular-polarization (CP) unit basis vectors,
$\hat e_\pm =(\hat x\pm i \hat y)/\sqrt{2}$.
The electric field radiated by the dipole at location $\vec r$ in the radiation zone has the form
\cite{Jackson99}
\be
\label{e3}
\vec E(\vec r,t)={k^2\over 4\pi\epsilon_0 r}[\vec p-(\vec p\cdot \hat n)\hat n]\,\exp(ikr-i\omega t),
\ee
where $k=\omega/c$ is the propagation constant of radiation and $\hat n=\vec r/r$ is the unit observation
vector. In view of (\ref{e2}) and since 
\be
\label{e4}
\hat n = {1\over \sqrt{2}}\sin\theta\left(e^{i\phi}\hat e_-+e^{-i\phi}\hat e_+\right) +\cos\theta \hat z, 
\ee
up to the linear order in the paraxial angle $\theta$, the electric field (\ref{e3}) takes the form
\ba
\label{e5}
\vec E(\vec r)={k^2\over 4\pi\epsilon_0 r}&\Biggl[p_+ \hat e_++p_-\hat e_-\nn
 &-{\theta\over \sqrt{2}}(p_+e^{i\phi} + p_- e^{-i\phi})\hat z\Biggr]\exp(ikr),
\end{align}
where we have omitted the time dependence of the field for the sake of brevity.
It is the longitudinal ($z$) component of the electric field in (\ref{e5}) that 
yields a helical component to the Poynting vector responsible for the AM of 
the radiation field.

%
In the thin-lens and paraxial (Fresnel) propagation limits that we assume here, in passing 
through the lens aperture with the spiral phase structure (\ref{e1}),
the transverse field, $\vec E_>^{(T)}(\vec r,t)$, acquires 
the phase shift, $-kR^2u^2/(2f)-\Psi(u,\phi)$,
where $f$ is the lens focal length. Just past the aperture, it thus has the form
\be
\label{e7}
\vec E_>^{(T)}(\vec r)={k^2\over 4\pi\epsilon_0 z_0}\left[p_+ \hat e_+ + p_-\hat e_-\right]
\exp[i\Phi(\vec u)],
\ee
where $r$ has been replaced by the axial source distance, $z_O$, from the lens pupil in the denominator
but by its more accurate, paraxial form, $z_O+R^2u^2/(2z_O)$, inside the phase factor $\exp(ikr)$ of (\ref{e5}).
As a result, the spatial phase function in the pupil has the form, 
\be
\label{e8}
\Phi(\vec u) = kz_O +{kR^2\over 2}\left({1\over z_O}-{1\over f}\right) u^2-\Psi(\vec u).
\ee

For lens apertures that are large compared to the wavelength, 
the transverse components of the electric field vector diffract approximately via
the scalar Fresnel-diffraction formula \cite{Goodman96}. As such, 
$\vec E^{(T)}(\vec r_I)$ at a distance $z_I$ from the lens aperture is the sum of its two CP components,
\be
\label{e8a}
\vec E^{(T)}(\vec r_I)=\vec E^{(T)}_+(\vec r_I) + \vec E^{(T)}_-(\vec r_I),
\ee
that may be expressed as
\ba
\label{e9}
\vec E^{(T)}_\pm(\vec r_I) =&{k^2R^2e^{ik(z_O+z_I)}\over 4\pi\epsilon_0z_O i\lambda z_I}p_\pm \hat e_\pm \nn
                              &\times\int P(u) e^{i\zeta u^2-i\Psi(\vec u)-i2\pi\vec u\cdot\vec s} d^2u.
\end{align}
Here $P(u)$ is the aperture function that is equal to 1 inside the aperture, {\it i.e.,}
for $u<1$, and 0 outside, and $\zeta$ is the defocus phase at the edge of the pupil, defined as 
\be
\label{e9a}
\zeta ={kR^2\over 2}\left({1\over z_O}+{1\over z_I}-{1\over f}\right)=
       {kR^2\over 2}{\delta z_O\over z_O(z_O+\delta z_O)},
\ee
where $\delta z_O$ is the distance of the source from the plane of best paraxial focus for which the thin-lens
equation holds.
In (\ref{e9}), we have scaled the transverse image-plane position vector, $\vec \rho_I$, by dividing it by 
the characteristic size of the Airy diffraction spot,
$\lambda z_I/R$, to arrive at $\vec s=\vec \rho_I /(\lambda z_I/R)$. 
For brevity we have suppressed here and in the expressions to follow 
the space-dependent phase factor $\exp[ik \rho_I^2/(2z_I)]$. 

A quarter-wave of defocus phase, {\it i.e.} $\zeta=\pm\pi/2$, was defined by Rayleigh \cite{Mahajan91}
to correspond to the characteristic depth of field (DOF)
for a clear-aperture imager. By contrast, an $L$-zone rotating-PSF imager 
has a DOF that corresponds
to $\pm L/2$ waves of defocus phase \cite{SP13}, which is $2L$ times as large as the Rayleigh DOF. 

From expression (\ref{e9}) for the transverse field components, we can construct the 
corresponding longitudinal 
components, $E^{(z)}_\pm$, in the image plane by imposing the transversality of the full field, 
$\vec\nabla\cdot\vec E=0$. For paraxial propagation, $\partial E^{(z)}_\pm/\partial z$ may
be replaced approximately by $ik E^{(z)}_\pm$, so the transversality condition is equivalent to
$ik E^{(z)}_\pm= -\vec \nabla^{(T)}\cdot\vec E^{(T)}_\pm$, which gives
\ba
\label{e10}
E^{(z)}_\pm(\vec s) = &{i\over k}\vec\nabla^{(T)}\cdot\vec E^{(T)}_\pm(\vec r_I)\nn
                    =&{-ik^3R^3e^{ik(z_0+z_I)}\over 8\pi^2\epsilon_0 z_0 z_I^2}p_\pm \nn
                      &\times \int P(u) \hat e_\pm\cdot\vec u \, e^{i\zeta u^2 -i\Psi(\vec u)-i2\pi\vec u
                       \cdot\vec s}d^2u\nn
                    =&{-ik^3R^3e^{ik(z_0+z_I)}\over 8\sqrt{2}\pi^2\epsilon_0 z_0 z_I^2}p_\pm \nn
                     &\times \int P(u)\, u\, e^{i\zeta u^2 -i\Psi(\vec u)\pm i\phi_u-i2\pi\vec u\cdot\vec s}d^2u,
\end{align}
where we used the identity $\hat e_\pm\cdot\vec u = u\exp(\pm i\phi_u)/\sqrt{2}$
to reach the final expression. We also ignored a negligibly small contribution to the transverse divergence 
in (\ref{e10}) from the phase factor $\exp[ik\rho_I^2/(2z_I)]$ that was omitted in (\ref{e9}). 
As expected, it is the longitudinal components of the electromagnetic field
that clearly exhibit the SAM for the two
CP components via the phase factors $\exp(\pm i\phi_u)$ in (\ref{e10}). 


From the first expression in (\ref{e10}), we expect that the longitudinal
electric field is of order $1/(k\lambda z_I/R)$, or order $R/z_I$, of the corresponding
transverse field in magnitude. As a result, a sensor pixel that is equally sensitive to 
all three components of the electric field of radiation impinging on it will see a contribution from the
spin-modified longitudinal component to the total counts that is of order $(R/z_I)^2$ of the counts contributed
by the transverse components of the field. For a high image-space NA, the SAM-dependent contribution can thus
be a significant fraction of the total PSF power and thus sensitively encode the polarimetric 
state of the source emission.


The lowest-order characteristics of source polarization are
completely determined by the bilinear statistical correlations of the two helicity components of the dipole
source, $p_\pm$. We assume, for definiteness, that it is the relative phase of the two
components, not their amplitudes, that is randomly distributed about a mean value, but taking a more
general statistical distribution of the two components would not change the expressions below materially. Writing 
\be
\label{e11}
p_\pm =r_\pm \exp(i\phi_\pm),
\ee
where the relative phase $\Delta\phi=\phi_+-\phi_-$ has the mean value $\phi_0$ and $r_\pm$ are the 
non-negative amplitudes of the two helicity components, 
we define the four Stokes parameters of emission, up to a scale factor, as
\ba
\label{e12}
s_0=&\langle |p_+|^2\rangle + \langle |p_-|^2\rangle = r_+^2 + r_-^2;\nn
s_1=&2{\rm Re} \langle p_+ p_-^* \rangle = 2\mu r_+r_-\cos\phi_0;\nn
s_2=&2{\rm Im} \langle p_+p_-^* \rangle= 2\mu r_+r_-\sin\phi_0;\ {\rm and}\nn
s_3=&\langle |p_+|^2\rangle - \langle |p_-|^2\rangle = r_+^2 - r_-^2;
\end{align}
where the triangular brackets denote expectation over the statistics of the relative phase. 
We have taken this distribution to be symmetric around the mean, which yields the 
expectation, $\langle \exp(i\Delta\phi)\rangle =\mu \exp(i\phi_0)$,
with $\mu$ being a real quantity of magnitude less than 1. 

The degree of polarization is the ratio
\be
\label{e14}
P={\sqrt{s_1^2+s_2^2+s_3^2}\over s_0},
\ee
which from definitions (\ref{e12}) is readily expressed in terms of $\mu$ and $r_\pm$ as
\be
\label{e15}
P=\sqrt{1-(1-\mu^2){4r_+^2r_-^2\over (r_+^2+r_-^2)^2}}.
\ee

For sensor pixels that detect the incident field components isotropically, the probability of photodetection 
is proportional to the expectation of the total time-averaged image-plane intensity, which is the 
squared modulus of the total image-plane field.
The expected time-averaged image intensity, $I(\vec r_I)$, may thus be expressed as
\ba
\label{e17}
I(\vec r_I) = &\langle |\vec E_+(\vec r_I) +\vec E_-(\vec r_I)|^2\rangle\nn 
            = &\langle |\vec E_+^{(T)}|^2\rangle+ \langle |\vec E_-^{(T)}|^2\rangle + 
              \langle |E_+^{(z)}+E_-^{(z)}|^2\rangle,
\end{align}
in which to arrive at the second line, we expressed the two helicity contributions to the
field in terms of their transverse and longitudinal components and then used the
identities, $\hat e_+\cdot\hat e_-^*=0$ and $\hat e_\pm\cdot\hat z=0$.
Substituting expressions (\ref{e9}) and (\ref{e10}) into (\ref{e17}) and using expressions 
(\ref{e12}) for the Stokes parameters, we may write $I(\vec r_I)$ as
\ba
\label{e18}
I(\vec r_I)= &
             \left({k^3R^2\over 8\pi^2\epsilon_0 z_Oz_I}\right)^2\left\{s_0\, |I^{(T)}|^2+{R^2\over 2z_I^2}
            \Bigl[{1\over 2}(s_0+s_3)|I^{(z)}_+|^2\right.\nn
            +{1\over 2}&\left. (s_0-s_3)|I^{(z)}_-|^2+
            \sqrt{s_1^2+s_2^2}{\rm Re}\left(I_+^{(z)}I_-^{(z)*}
            e^{i\phi_0}\right)\Bigr]\right\},
\end{align}
where $I^{(T)}$ and $I^{(z)}_\pm$ are defined as the integrals
\ba
\label{e19}
I^{(T)}(\vec s)=&\int P(u)\, e^{i\zeta u^2 -i\Psi(\vec u)-i2\pi\vec u\cdot\vec s}d^2u;\nn
I^{(z)}_\pm(\vec s)=&\int P(u)\, u\, e^{i\zeta u^2 -i\Psi(\vec u)\pm i\phi_u-i2\pi\vec u\cdot\vec s}d^2u.
\end{align}

For the spiral phase distribution (\ref{e1}) with $f_l(\phi_u) =l\phi_u$, the angular integrations may
be performed exactly over the different annular zones, and the integrals (\ref{e19}) reduce
to the following sums of radial integrals:
\ba
\label{e20}
I^{(T)}(\vec s)=&2\pi\sum_{l=1}^L (-i)^le^{-il\phi}\int_{\sqrt{(l-1)/L}}^{\sqrt{l/L}} e^{i\zeta u^2} J_l(2\pi u s)\, u\, du;\nn
I^{(z)}_\pm(\vec s)=&2\pi\sum_{l=1}^L (-i)^{l\mp 1}e^{-i(l\mp 1)\phi}\int_{\sqrt{(l-1)/L}}^{\sqrt{l/L}} 
e^{i\zeta u^2} \nn
&\qquad \qquad \times J_{l\mp1}(2\pi us)\, u^2 du.
\end{align}
These radial integrals and thus (\ref{e20}) may be evaluated numerically and the result
substituted into (\ref{e18}) to determine the full PSF for an arbitrary polarization state 
of the source dipole. 

For $s,\zeta<<L$, each of the $u$ integrals over the $l$th zone in expression (\ref{e20}) 
may be well approximated by the radial zone width times the integrand at the mid 
point of the integration range.
This yields a $\zeta$ dependence of the integrals as $\exp[i\zeta(l-1/2)/L]$
which when combined with the $\exp[-i(l\mp 1)\phi]$ prefactor in the $l$th term of each expression in (\ref{e20})
confirms the spatial rotation of the magnitudes of $I^{(T)}$ and $I_\pm^{(z)}$ at a uniform rate with 
changing $\zeta$, but
the phases of $I_\pm^{(z)}$, unlike that of $I^{(T)}$, are clearly seen to have different
residual contributions, $\pm\phi$, for the two 
different photon helicities. This means that in (\ref{e18})
the last term, which encodes $s_1$ and $s_2$, does not rigidly rotate with changing $\zeta$,
while the other terms involving only the magnitudes of the integrals (\ref{e19}) do so uniformly without 
change of shape or size. Thus when $s_1$ and $s_2$ are significantly different from zero,
the polarization encoding for a high-NA imager is attended by a compromised rotational character of the PSF power.

The results of numerical evaluation of (\ref{e18}) are
displayed in Fig.~1 for four different polarization states of the source,
specifically the unpolarized state, the helicity $\pm 1$ states, and the $x$-polarized state, 
for two different axial depths, $\zeta=0$ and 8, and for $L=7$ zones in the spiral phase mask. The corresponding 
Stokes vectors are proportional to $(1,0,0,0)$, $(1,0,0,\pm1)$, and $(1,1,0,0)$, respectively. The image-space NA 
was chosen to be large \cite{FN1} at $R/z_I=1$.
\begin{figure}[htbp]
\hspace{.001cm}
\subfloat[]
{\includegraphics[width=.75in, height=.75in]{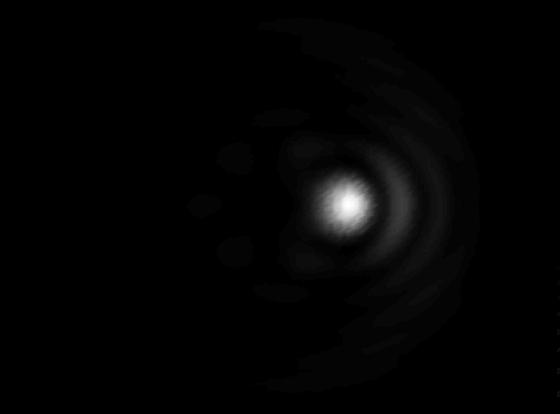}}
\hspace{.001cm}
\subfloat[]
{\includegraphics[width=.75in, height=.75in]{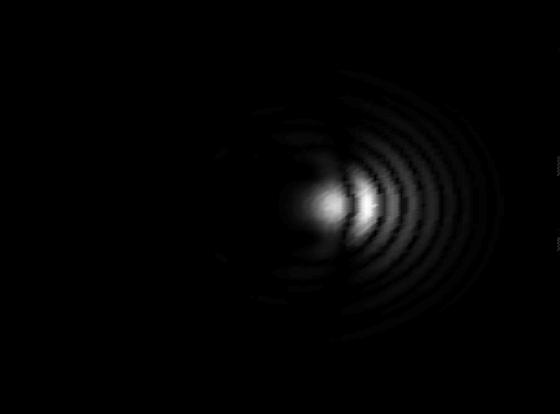}}
\hspace{.001cm}
\subfloat[]
{\includegraphics[width=.75in, height=.75in]{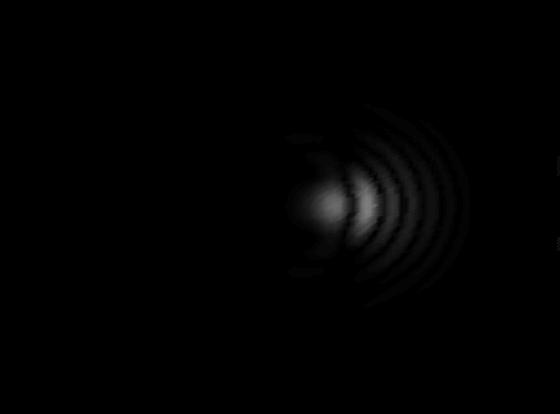}}
\hspace{.001cm}
\subfloat[]
{\includegraphics[width=.75in, height=.75in]{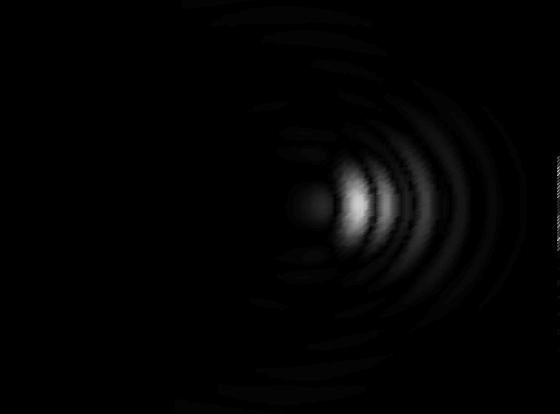}}\\
\hspace{.001cm}
\subfloat[]
{\includegraphics[width=.75in, height=.75in]{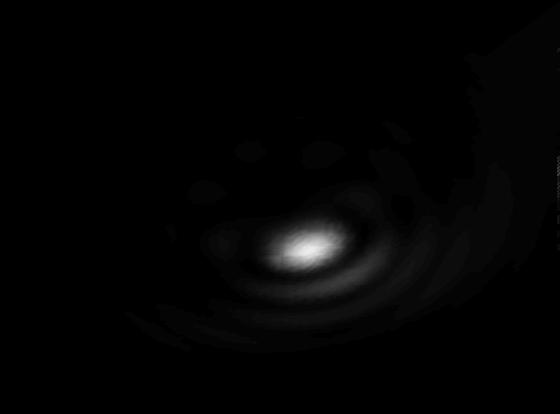}}
\hspace{.001cm}
\subfloat[]
{\includegraphics[width=.75in, height=.75in]{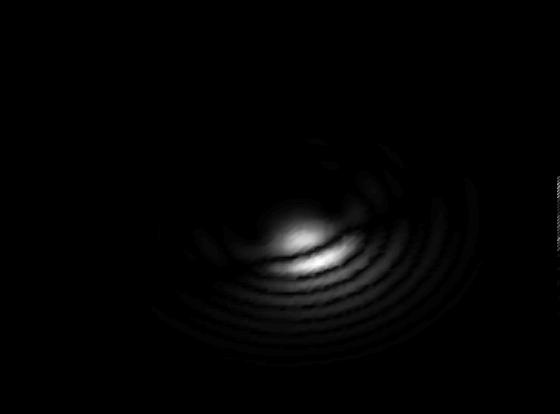}}
\hspace{.001cm}
\subfloat[]
{\includegraphics[width=.75in, height=.75in]{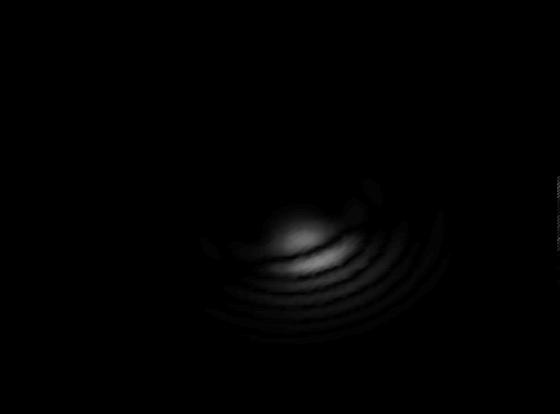}}
\hspace{.001cm}
\subfloat[]
{\includegraphics[width=.75in, height=.75in]{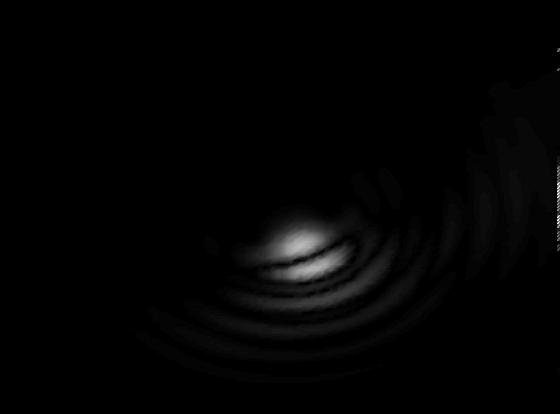}}
\vspace{-10pt}
\caption{PSF for a helicity +1 source 
at (a) zero defocus and (e) defocus 8. The plots (b)-(d) display 
the magnitude of the difference between the PSF power when the source is in helicity -1, unpolarized, 
and $x$-polarized states, respectively, and the PSF power shown in (a) for the helicity +1 source, all
for zero defocus. Plots (f)-(h) are the corresponding difference-magnitude plots for defocus 8.
For the ease of visualization, the gray scale used for the difference plots is 1/6.8 times that used 
for the full PSF plots (a) and (e).}
\vspace{-10pt}
\end{figure}  
The 10-15\% variation in the PSF (\ref{e18}) with changing polarization seen in these figures should be sufficient 
to recover sensitively the polarization state of the source emission in an imager of sufficiently high NA, 
as we next confirm by computer simulation. 

We simulated image data for a single point-dipole emitter operating under varying SNR conditions for the
Gaussian additive-sensor-noise and Poisson photon-shot-noise models. Three different closely
spaced values of the Stokes vector were selected for the simulation, all with $s_0=1$, $s_1=s_2=0$, but with 
$s_3$ taking values 0.8, 0.9, and 1. The first two values of $s_3$ correspond to the source 
polarization being in a mixed state in which a small incoherent admixture of the negative-helicity 
CP state corrupts the purity of the positive-helicity CP state, while the last value of $s_3$ represents 
the pure positive-helicity CP state. 
We formulate the inverse problem of reconstructing the Stokes vector from the noisy image data as 
a simple $\chi^2$-minimization problem with respect to the three free Stokes parameters, $s_1$, $s_2$, and $s_3$,
with the value of $s_0$ fixed (here at 1) as the overall photon-flux normalization parameter.  We employed
the Matlab code {\it fminunc} to perform the minimization for each noisy image realization
at each value of the peak SNR (PSNR), defined as the ratio of the peak image-pixel signal value, $I_0$, and the 
standard deviation of the noise at that pixel, the latter being equal to
$I_0^{1/2}$ for the shot-noise case. The ratio $R/z_I$ was again chosen to be 1.
\begin{figure}
\subfloat[]
{\includegraphics[width=1.8in, height=1.75in]{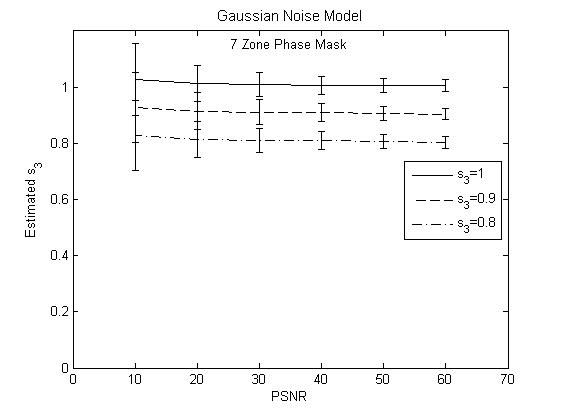}}
\subfloat[]
{\includegraphics[width=1.8in, height=1.75in]{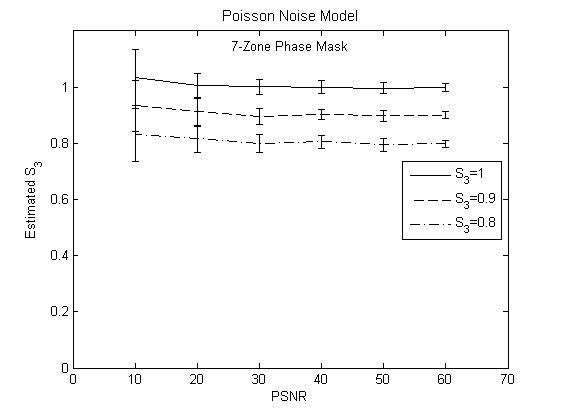}}
\vspace{-10pt}
\caption{Reconstructed Stokes parameter $s_3$, shown with $\pm 1\sigma$ error bars, vs. PSNR for the (a) Gaussian
additive sensor noise and (b) Poisson shot noise models.}
\vspace{-18pt}
\end{figure}

From the plots of the reconstructed $s_3$, with $\pm 1\sigma$ error bars obtained from reconstructions
from 30 different noise realizations for each PSNR value in each noise model, 
we see that at PSNR values exceeding 30, 
we can discriminate between closely spaced values of the $s_3$ parameter at the 10\% level from the 
polarization-dependent PSFs of the kind shown in Fig.~1. The performance under the Poisson noise
model is somewhat superior, with smaller error bars. For either noise model, higher statistical fidelity of 
discrimination than at the $\pm 1\sigma$ level would obviously require higher PSNR values.
Although not shown here, similar error bars
were obtained for other choices of the Stokes vector too, confirming its robust recovery by
our polarimetric imager under fairly general conditions.
Finally, we simulated image data for spiral phase masks with different total zone numbers and verified that, as expected, with 
a smaller (larger) number of zones in the mask, the relative contribution of the SAM-dependent wave polarization 
to PSF rotation is larger (smaller), thus lowering (raising) the PSNR threshold for a reliable estimation 
of the Stokes vector.

Unlike other polarimetric imagers \cite{GPY10,Sasagawa13}, our rotating-PSF-based polarimetric imager can
sense both the 3D locations and polarization states of point sources in a single snapshot over 
a large focal volume without requiring specialized sensing elements.  
The performance of the proposed imager for multiple, closely spaced point sources in a 3D scene
will be treated elsewhere.

The work was supported by the US Air Force Office of Scientific Research under grant no. F9550-11-1-0194.


\begin{thebibliography}{99}
\bibitem{FN1} A large NA is not entirely consistent with paraxial propagation assumed here, but is used
here mainly to illustrate its importance for polarimetric imaging.
\end{thebibliography}
\end{document}